\documentclass[aps,prmaterials,superscriptaddress,reprint,amsmath,amssymb,longbibliography,floatfix]{revtex4-2}

\usepackage[utf8]{inputenc}
\usepackage[T1]{fontenc}
\usepackage{amsmath}
\usepackage{amssymb}
\usepackage{graphicx}
\usepackage{epstopdf}
\usepackage{color}
\usepackage{enumerate}
\usepackage{ulem}
\usepackage{multirow}
\usepackage{array}
\usepackage{float}
\usepackage{amsmath}
\usepackage{hyperref}
\usepackage{epsfig}
\usepackage{bm}
\usepackage{hyphenat}
\usepackage{makecell}
\usepackage{color,soul}
\usepackage[dvipsnames]{xcolor}
\usepackage{chemformula}
\usepackage{siunitx}
\usepackage{xcolor}

\DeclareSIUnit\angstrom{\text {Å}}

\begin{document}

\title{Electrical and thermal magnetotransport in altermagnetic CrSb}

\author{Sajal Naduvile Thadathil}
\thanks{These authors contributed equally to this work.}
\affiliation{Hochfeld-Magnetlabor Dresden (HLD-EMFL) and W\"urzburg-Dresden
Cluster of Excellence ctd.qmat, Helmholtz-Zentrum
Dresden-Rossendorf, 01328 Dresden, Germany}
\affiliation{Institute of Solid State and Materials Physics and W\"urzburg-Dresden Cluster of Excellence ctd.qmat, Technische Universität Dresden, 01062 Dresden, Germany}

\author{Christoph M\"uller}
\thanks{These authors contributed equally to this work.}
\affiliation{Institute of Physics, Czech Academy of Sciences, Cukrovarnická 10, 162 00 Prague 6, Czech Republic}
\affiliation{Faculty of Mathematics and Physics, Charles University, 12116 Prague, Czech Republic}

\author{Reza Firouzmandi}
\affiliation{Leibniz Institute for Solid State and Materials Research, IFW Dresden, Helmholtzstr. 20, 01069 Dresden, Germany}

\author{Lorenz Farin}
\affiliation{Hochfeld-Magnetlabor Dresden (HLD-EMFL) and W\"urzburg-Dresden
Cluster of Excellence ctd.qmat, Helmholtz-Zentrum
Dresden-Rossendorf, 01328 Dresden, Germany}
\affiliation{Institute of Solid State and Materials Physics and W\"urzburg-Dresden Cluster of Excellence ctd.qmat, Technische Universität Dresden, 01062 Dresden, Germany}

\author{Srikanta Goswami}
\affiliation{Institute of Physics, Czech Academy of Sciences, Cukrovarnická 10, 162 00 Prague 6, Czech Republic}

\author{Antonin Badura}
\affiliation{Institute of Physics, Czech Academy of Sciences, Cukrovarnická 10, 162 00 Prague 6, Czech Republic}

\author{Pascal Manuel}
\affiliation{ISIS Neutron and Muon Source, STFC Rutherford Appleton Laboratory, Harwell campus, Didcot OX11 0QX, Oxfordshire, United Kingdom}

\author{Fabio Orlandi}
\affiliation{ISIS Neutron and Muon Source, STFC Rutherford Appleton Laboratory, Harwell campus, Didcot OX11 0QX, Oxfordshire, United Kingdom}

\author{Philipp Ritzinger}
\affiliation{Institute of Physics, Czech Academy of Sciences, Cukrovarnická 10, 162 00 Prague 6, Czech Republic}
\affiliation{Faculty of Mathematics and Physics, Charles University, 12116 Prague, Czech Republic}

\author{Václav Petříček}
\affiliation{Institute of Physics, Czech Academy of Sciences, Cukrovarnická 10, 162 00 Prague 6, Czech Republic}

\author{Marc Uhlarz}
\affiliation{Hochfeld-Magnetlabor Dresden (HLD-EMFL) and W\"urzburg-Dresden
Cluster of Excellence ctd.qmat, Helmholtz-Zentrum
Dresden-Rossendorf, 01328 Dresden, Germany}

\author{Tommy Kotte}
\affiliation{Hochfeld-Magnetlabor Dresden (HLD-EMFL) and W\"urzburg-Dresden
Cluster of Excellence ctd.qmat, Helmholtz-Zentrum
Dresden-Rossendorf, 01328 Dresden, Germany}

\author{Michal Baj}
\affiliation{Faculty of Physics, University of Warsaw, Pasteura 5, Warsaw, Poland}

\author{Marein C. Rahn}
\affiliation{Institute of Solid State and Materials Physics and W\"urzburg-Dresden Cluster of Excellence ctd.qmat, Technische Universität Dresden, 01062 Dresden, Germany}

\author{Thanassis Speliotis}
\affiliation{Institute of Nanoscience and Nanotechnology, National Center for Scientific Research Demokritos, 15341 Athens, Greece}

\author{Markéta Žáčková}
\affiliation{Faculty of Mathematics and Physics, Charles University, 12116 Prague, Czech Republic}

\author{Jiří Pospíšil}
\affiliation{Faculty of Mathematics and Physics, Charles University, 12116 Prague, Czech Republic}

\author{Bernd~B\"uchner}
\affiliation{Institut f\"ur Festk\"orperforschung, Leibniz IFW Dresden, 01069 Dresden, Germany}
\affiliation{Institute of Solid State and Materials Physics and W\"urzburg-Dresden Cluster of Excellence ctd.qmat, Technische Universität Dresden, 01062 Dresden, Germany}
\affiliation{Center for Transport and Devices, Technische Universität Dresden, 01069 Dresden, Germany}

\author{Jochen Wosnitza}
\affiliation{Hochfeld-Magnetlabor Dresden (HLD-EMFL) and W\"urzburg-Dresden
Cluster of Excellence ctd.qmat, Helmholtz-Zentrum
Dresden-Rossendorf, 01328 Dresden, Germany}
\affiliation{Institute of Solid State and Materials Physics and W\"urzburg-Dresden Cluster of Excellence ctd.qmat, Technische Universität Dresden, 01062 Dresden, Germany}

\author{Helena Reichlov\'a}
\affiliation{Institute of Physics, Czech Academy of Sciences, Cukrovarnická 10, 162 00 Prague 6, Czech Republic}

\author{Vilmos Kocsis}
\email{v.kocsis@ifw-dresden.de}
\affiliation{Leibniz Institute for Solid State and Materials Research, IFW Dresden, Helmholtzstr. 20, 01069 Dresden, Germany}

\author{Toni Helm}
\email{t.helm@hzdr.de}
\affiliation{Hochfeld-Magnetlabor Dresden (HLD-EMFL) and W\"urzburg-Dresden
Cluster of Excellence ctd.qmat, Helmholtz-Zentrum
Dresden-Rossendorf, 01328 Dresden, Germany}

\author{Dominik Kriegner}
\email{kriegner@fzu.cz}
\affiliation{Institute of Physics, Czech Academy of Sciences, Cukrovarnická 10, 162 00 Prague 6, Czech Republic}

\date{\today}

\begin{abstract}
Chromium antimonide has emerged as a key material platform for studying altermagnetism because of its simple binary composition, high Néel temperature, and semimetallic electronic structure.
Here, we investigate electrical and thermal magnetotransport in single-crystalline CrSb using steady-and pulsed-magnetic fields up to 65\,T, and complement these measurements with neutron diffraction and magnetization data.
We confirm the compensated magnetic structure and observe a large nonsaturating magnetoresistance together with a pronounced nonlinear Hall response at low temperatures.
Multicarrier modeling, supported by mobility-spectrum analysis, reveals coexisting electron- and hole-like charge carriers with mobilities up to $\sim 3000$\,cm$^{2}$/Vs and shows that the number of transport channels that can be resolved strongly depends on the accessible magnetic-field range.
Thermal-transport measurements further reveal a nonlinear thermal Hall response and a thermal conductivity substantially exceeding a simple Wiedemann–Franz law.
The broadly similar field and temperature evolution of electrical and thermal transport point to a dominant electronic contribution, while the remaining deviations indicate additional heat-carrying channels.

\end{abstract}

\maketitle
\section{Introduction}
    
Recent theoretical predictions and experimental observations on magnetic materials have significantly advanced with the identification of a new phase of magnetism known as altermagnetism~\cite{Smejkal2022,Smejkal2022b,Mazin2022}.
The key signature that sets altermagnets (AMs) apart from conventional collinear antiferromagnets is the presence of strong nonrelativistic spin splitting in its electronic band structure despite an overall zero net magnetization~\cite{krempasky2024,lee2024,osumi2024,reimers2024,fedchenko2024, Jiang2025}.
This spin polarization alternates in sign for different momentum directions and gives rise to intriguing effects, such as spin-current generation~\cite{bai2022,bose2022}, giant magnetoresistance~\cite{Smejkal2022GMR}, and a spontaneous anomalous Hall effect (AHE)~\cite{gonzalez2023,wang2023, Kluczyk2024, Reichlova2024, Takagi2025}.

Of the many theoretically predicted candidate AMs~\cite{Smejkal2022, Guo2023, Gao2025}, various have been experimentally studied.
For example, \ch{RuO2}~\cite{Feng2022,wang2023,tschirner2023,bai2022,bose2022} and \ch{KV2Se2O}~\cite{Jiang2025} appeared to be very promising.
However, some of the results are currently heavily debated due to fragility or even absence of magnetic order in bulk RuO$_2$~\cite{Smolyanyuk2024, Kesler2024, Kiefer2025, Occhialini2026} as well as an incompatible magnetic structure found in \ch{KV2Se2O}~\cite{Sun2025}.
More recently, \ch{Mn5Si3} came into focus as a candidate AM~\cite{Gao2025, Leivisk2024,Rial2024,Han2024,Badura2025}, exhibiting a spontaneous anomalous Hall effect that is consistent with a $d$-wave character of its magnetic order~\cite{REGUS2012,Reichlova2024} and corresponding spin transport~\cite{mencos2025}.
Experimentally well established are the results on compounds adopting the NiAs structure (P6$_3$/mmc space group) such as MnTe~\cite{lee2024,osumi2024,aoyama2024,krempasky2024,gonzalez2023}, CrSb~\cite{reimers2024, urata2024, Bommanaboyena2025}, as well as the slightly distorted FeS~\cite{Takagi2025}.

Among these, CrSb stands out due to its high N\'eel temperature of around 700\,K~\cite{Takei1963,Snow1952}, and a significant spin-splitting energy of up to 1.2\,eV~\cite{Smejkal2022,Guo2023}.
It is a semimetal with a magnetic order that consists of collinear antiparallel spins aligned along the $c$ axis characterized by $^26 / ^2m ^2m ^1m$ spin point group, making it a $g$-wave AM. 
In this configuration, the magnetic sublattices are linked by mirror and rotational symmetries, which allows for a substantial spin splitting. 
Recent experimental studies using angle-resolved photoemission spectroscopy (ARPES) and soft x-ray measurements confirm the altermagnetic band splitting of CrSb~\cite{reimers2024, Yang2025}, with splitting energies reaching 0.6\,eV~\cite{reimers2024}.
Moreover, further ARPES experiments revealed the presence of surface Fermi arcs near the Fermi level for the (100) cleavage plane, associated with bulk topological Weyl nodes~\cite{Lu2025,Li2025}. 
These arcs are consistent with theoretical predictions and confirm the existence of topologically nontrivial surface states in CrSb and a coexistence of opposite-spin and same-spin Weyl points~\cite{Li2025}. 
These unique topological properties arise from the altermagnetic nature of CrSb and renders it a promising platform to study distinct electronic topologies with dominating non-relativistic band splitting.

While various AMs, in agreement with predictions, show a spontaneous AHE~\cite{Smejkal2022,gonzalez2023,Reichlova2024,Takagi2025}, the magnetic structure of CrSb does not allow for this to emerge~\cite{urata2024, Bommanaboyena2025}.
Nonetheless, a pronounced nonlinear Hall effect was observed in bulk CrSb~\cite{urata2024,bai2025,peng2024}.
The different studies, however, identified a varying number of bands that contribute to the electrical transport providing a possible explanation for the nonlinear Hall effect observed in CrSb.

Apart from electrical magnetotransport measurements, recent thermoelectric measurements on CrSb have revealed a significant conventional Seebeck response accompanied by an anomalous Nernst effect~\cite{Li2025Nernst}. This was attributed to the altermagnetic band structure near the Fermi level and supported by first-principles calculations. 
Beyond CrSb, theoretical works on AMs indicate pronounced anisotropies in the thermal conductivity, as shown in \ch{RuO2}~\cite{Zhou2024} and \ch{MnTe}~\cite{Liu2025}. These studies also propose thermal Hall and Nernst effects in suitable metallic and insulating altermagnets~\cite{Hoyer2025}. All of these phenomena arise from magnon Berry curvatures in insulating systems and electronic Berry curvature in metallic systems, provided the relevant altermagnetic symmetries are present.

In this study, we confirm the magnetic structure in single crystals of CrSb by neutron diffraction and magnetization measurements.
We present magnetotransport measurements conducted over an extended magnetic field range up to 65\,T.
We observe a non-saturating magnetoresistance (MR) and a nonlinear Hall resistance with magnetic field.
Our multi-carrier analysis of the transport data is consistent with the coexistence of high-mobility electron- and hole-type charge carriers.

Additionally, we conducted thermal-conductivity and thermal-Hall-effect (ThHE) measurements.
The ThHE signal exhibits a nonlinear field dependence that, similar to the electrical Hall effect, becomes more pronounced at low temperatures.
We further observe a marked deviation from a simple Wiedemann–Franz law.

This manuscript is structured into three main sections: Experimental Methods, Results and Discussion, where the findings are presented and analyzed in detail, and Conclusions. The Results and Discussion section is further divided into three parts: A. Growth, Structure, and Magnetism; B. Magnetoresistance and Hall Effect; and C. Thermal Conductivity and Thermal Hall effect.

\section{Experimental methods}

\subsection{Growth and structural characterization}

We used chemical vapor transport for the synthesis of high-quality CrSb single crystals. 
These crystals originate from the same batch as those used in the earlier XMCD study~\cite{Biniskos2025}. 
We cut samples for various measurements from these single crystals using a wire saw for bulk samples and gallium focused-ion-beam lithography for microstructures (see Appendix~\ref{app:FIB} for details).

We carried out structural characterization by Laue backscatter diffraction using a Photonic Science system.
Powder x-ray diffraction was performed using a Panalytical Empyrean system with Cu $K_\alpha$ radiation and a rotating sample at room temperature.
We conducted powder diffraction simulations with x-ray utilities~\cite{Kriegner:rg5038}.

\subsection{Neutron diffraction}

We performed neutron diffraction from a CrSb single crystal with the time-of-flight neutron diffractometer WISH (ISIS, Rutherford Appleton Laboratory)~\cite{Chapon2011}. 
A single crystal with volume of roughly \SI{15}{\milli \meter^3}, depicted in Fig.~\ref{fig1}(a), was mounted on an aluminium strip using aluminium tape for measurements conducted at various sample orientations at a temperature of 1.5\,K.
We processed the data using the Mantid~\cite{Arnold2014} and nuclear and magnetic structure refinement were performed within JANA2020 \cite{petricek2023, Henriques2024}.
The refinement was performed using 258 reflections, of which 256 were observed with finite intensity.
In addition to the lattice-parameter refinements during the data processing using Mantid, four parameters were refined. These were: (i) an intensity scaling factor (ii) the magnetic moment on the Cr site, (iii) isotropic atomic-displacement parameters (ADPs) which were constrained to be equal for both atomic species and (iv) extinction which was treated using the isotropic Becker–Coppens Type-2 model~\cite{Becker:a10603}.
\subsection{ Calorimetry and magnetometry}

We performed differential scanning calorimetry (DSC) on CrSb crystals using a SETSYS Evolution 24 instrument (SETARAM Instrumentation) with an S-type DSC plate rod easy fit (50 °C to 1500 °C). 
Standard alumina crucibles with a diameter of 5 mm and a height of 8 mm were used; the total mass of each sample was less than 10 mg. The temperature dependence of the heat flow was measured in an He atmosphere in the temperature range of 300–1000 K, with a heating/cooling rate of 5 K/min. Transition temperatures were determined as the onset of the observed peaks.

We performed magnetization measurements in a Quantum Design MPMS3 - SQUID magnetometer and a vibrating-sample magnetometer equipped with an oven insert for temperatures up to \SI{1000}{\kelvin}.

\subsection{Electrical magnetotransport}

We carried out high-temperature electrical resistivity in a rapid-thermal-annealing system, which can sweep temperature from 300 to \SI{850}{\kelvin}.
We used a bulk bar-shaped sample labeled S1 [see inset Fig.~\ref{fig2}(a)], with dimensions $(3\times 0.5\times 0.1)$\unit{\milli\meter^3}, cut from CrSb single crystals.
We used an electrical current of \SI{5}{\milli\ampere} at \SI{333.33}{\hertz}.

We did steady-field magnetotransport measurements on sample S2 [Fig.~\ref{fig3}(a)] using a 14\,T Physical Property Measuring System (PPMS). 
We performed resistance measurements using a standard AC four-point lock-in technique. 
We applied an electrical current of \SI{5}{\milli\ampere} at \SI{177.77}{\hertz} along specified crystallographic directions.

We performed pulsed-field magnetotransport measurements on sample labeled S2 as well as on a focused-ion-beam microfabricated sample labeled L1 [Fig.~\ref{fig3}(a)] in a 70\,T pulsed magnet with a pulse duration of 150\,ms at the Dresden High Magnetic Field Laboratory, Germany.
We recorded data for excitation currents of \SI{5}{\milli\ampere} at $30\,\mathrm{kHz}$ for S2, and \SI{100}{\micro\ampere} at $33.33\,\mathrm{kHz}$ for L1.
We applied a digital lock-in analysis after recording the raw AC voltages during each pulse.

\subsection{Thermal magnetotransport}

We carried out steady-field thermal conduction and thermal Hall effect on the bar-shaped sample labeled S3 [Fig.~\ref{fig:thermal}(a)] with dimensions $(6.7\times 1.5\times 0.29)$\unit{\milli\meter^3}, which was mounted in a six-point measurement geometry.
We measured thermal conductivity ($\kappa_\mathrm{xx}$) and the thermal Hall ($\kappa_\mathrm{xy}$) effects simultaneously. 
In our experiment, we applied the heat current within the hexagonal plane, while the magnetic field was applied out-of-plane ($\mu_0H\parallel c$). 
We used a \SI{2.7}{\kilo\ohm} chip resistor as a heat source and measured the input power using a four-point method. 
Both the longitudinal ($l_{xx}$ = 3.26 mm) and transversal ($l_{xy}$ = 1.5 mm) thermal gradients were measured with field-calibrated 2-mil (\SI{50}{\micro\meter}) E-type thermocouple pairs (Goodfellow Advanced Materials). 
For the sensitive measurement of $\kappa_\mathrm{xx}$ and $\kappa_\mathrm{xy}$, we used a steady-state differential method: We measured the difference in the thermocouple signals in the presence and absence of a constant heat current at a stable base temperature and magnetic field.

\section{Results and Discussion}

\subsection{Growth, Structure and Magnetism}

\begin{figure}[tb]
    \centering
    \includegraphics[
    width=0.9\columnwidth] {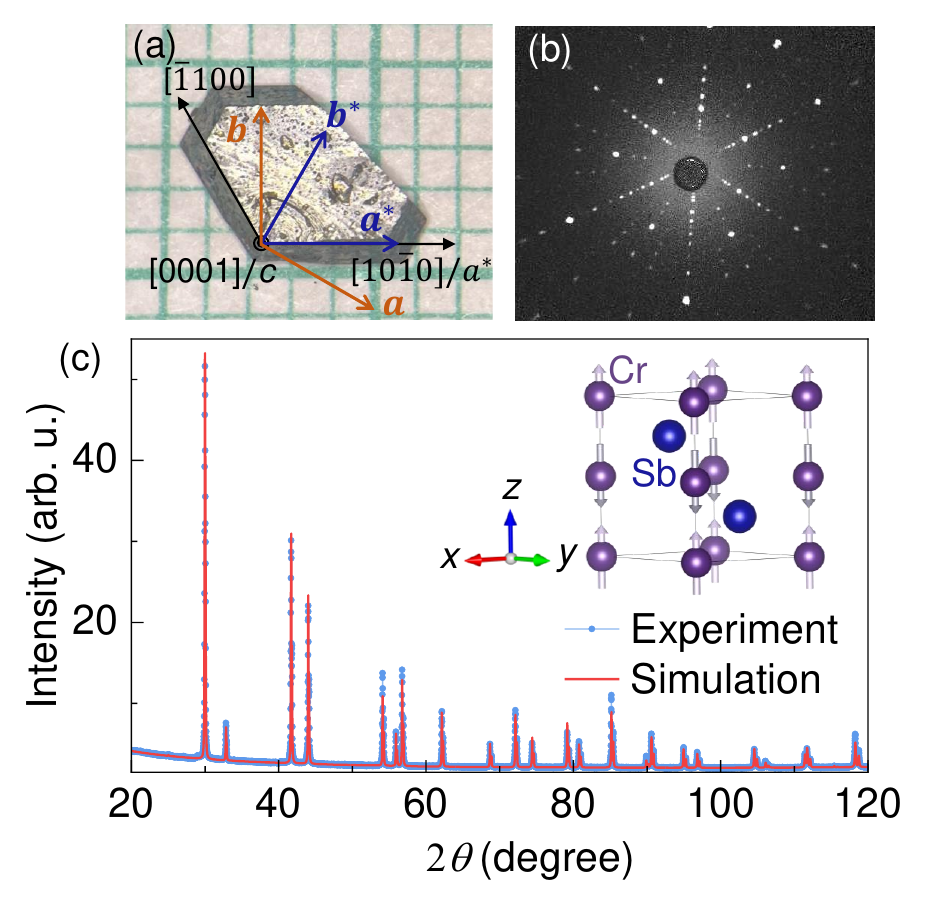} 

    \caption{\textbf{Characterization of single crystals of CrSb: (a)}
    Optical image of bulk CrSb single crystal with [0001]/$c$ axis perpendicular to the surface.
   The axes $[10\bar10]$/$a^{*}$,[0001]/$c$ and $[\bar1100]$ are indicated. 
Blue axes indicate reciprocal lattice directions, whereas orange indicates real space directions.
    \textbf{(b)} Room temperature backscattered Laue diffraction pattern obtained from a flat single crystal. 
    \textbf{(c)} Room-temperature powder x-ray diffraction pattern of CrSb.
    The inset shows the hexagonal unit cell of CrSb together with its magnetic structure (visualized using \textsc{vesta}~\cite{Momma2011}).
    The magnetic moments of the chromium atoms are directed along the $c$ axis.
    }
    \label{fig1} 

\end{figure}

We used chemical vapor transport for the growth of \unit{\milli\meter}-sized single crystals [Fig.~\ref{fig1}(a)].
The obtained crystals are typically plate-like with the $c$ axis perpendicular to the largest facet.
We confirmed the crystalline quality by Laue diffraction patterns as shown in Fig.~\ref{fig1}(b).
The Laue pattern exhibits a pronounced sixfold rotational symmetry, consistent with the hexagonal $c$ axis oriented perpendicular to the platelet surface.
Repeated measurements at multiple positions on the same crystal yielded consistent results, demonstrating the uniform crystallographic orientation across the probed area.

Given that Laue diffraction is limited by the short penetration depth of x-rays, we also performed x-ray diffraction measurements on powder prepared from multiple grown crystals to ensure bulk phase purity.
The obtained diffraction pattern [Fig.~\ref{fig1}(c)] can be described by a nontextured model of single-phase CrSb, which verifies the phase purity, i.e., confirming the exclusive presence of CrSb in the NiAs phase.
The determined lattice constants, $a = \SI{4.122(2)}{\angstrom}$ and $c = \SI{5.469(2)}{\angstrom}$, agree with those reported previously~\cite{Snow1952,REGUS2012}. 

\begin{figure}[tb]
    \centering
    \includegraphics[width=1\columnwidth] {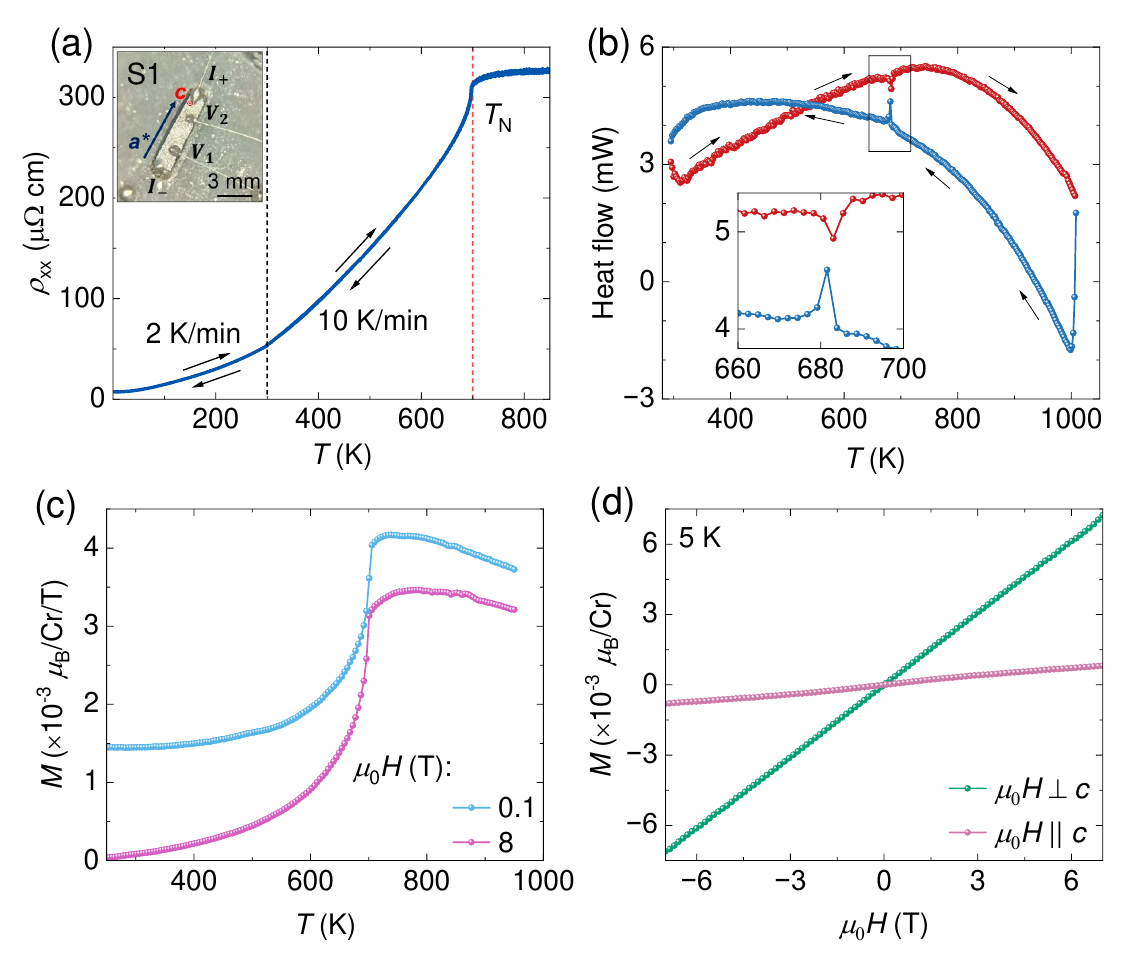} 
    
    \caption{\textbf{Determination of the N\'eel temperature: (a)} Zero-field resistance from 300 to 850\,K of a bulk CrSb single crystal (inset), showing $T_\mathrm{N} \approx 700$ K.
    \textbf{(b)} Differential scanning calorimetry thermograms of CrSb from \SI{300}{\kelvin} to \SI{1000}{\kelvin}.
    \textbf{(c)} Magnetization of CrSb measured from room temperature to 1000 K with magnetic fields of 0.1 T and 8 T, indicating a N\'eel temperature of about 700\,K.
    \textbf{(d)} Magnetization at 5\,K for magnetic fields applied along $c$ as well as perpendicular to c within the basal plane.
    }
    \label{fig2}
\end{figure}

The onset of magnetic order in CrSb is detectable in various properties.
For example, a notable transition in the resistivity is observed around 697\,K (estimated by the second derivative) in a bulk sample, as depicted in Fig.~\ref{fig2}(a) for both heating and cooling.
Above this temperature, the resistivity saturates.
We conducted differential scanning calorimetry, as shown in Fig.~\ref{fig2}(b), and found a clear heat-flow peak [inset of Fig.~\ref{fig2}(b)] indicating the magnetic transition at $T_\mathrm{N} = (680\pm5)$\,K.
The discrepancy to previously reported N\'eel temperature ($T_\mathrm{N}\approx713$\, K)~\cite{Toshiro1957, Hirone1956} as well as between the different measurements may arise from a different temperature measurement schemes used by different probes.
The magnetic origin of the observed anomalies is confirmed by the temperature-dependent magnetometry measurements shown in Fig.~\ref{fig2}(c).
The magnetization reveals a clear transition to a magnetically compensated state, whose transition temperature is largely independent of the applied external magnetic field up to 8\,T.
Even for a magnetic field of \SI{1}{\tesla}, we detect a magnetization of only a few $10^{-3}$\,$\mu_\mathrm{B}$ / Cr / T
The magnetization recorded during magnetic field sweeps, confirms this picture and reveals no spontaneous uncompensated magnetic moment [Fig.~\ref{fig2}(d)].
The susceptibility of CrSb is, however, found to be larger for $\mu_0H\perp c$ as compared to $\mu_0H\parallel c$.
This indicates a magnetic anisotropy with magnetic moments predominantly along the direction with low susceptibility, that is along the $c$ axis.

To confirm the magnetic structure and, specifically, the N\'{e}el-vector orientation, we performed neutron diffraction on a single crystal.
Our analysis focused on a potential lowering of the magnetic symmetry from the previously reported magnetic space group (MSG) $P6_{3}'/m'm'c$ \cite{Takei1963,Snow1952} (corresponding to the irreducible representation $m\Gamma_4^+$ with $^26 / ^2m ^2m ^1m$ spin point group).
Such a symmetry reduction, specifically a rotation of the N\'{e}el vector, was predicted to induce an anomalous Hall effect \cite{Yu2025}.
However, our refinements, covering all possible MSGs obtained from the representation analysis performed using JANA2020 and their subgroups, indicate that no other compensated magnetic order is compatible with the observed diffraction data.
The other MSGs tested are: $P6_{3}/mm'c'$, $P6_{3}'/2'2$, $P6_{3}'$, $Am'm'2$, $Cmcm$, $Cm'c'm$, $Cmc'm'$, $Cm'cm'$, $P\bar{3}1c$, $P\bar{1}$, $P2_{1}'$, $P2_{1}'/m'$, $P2_{1}/m$, $P{3}$, $C{c}$, $C{2'}$, $C2/c$, $C2'/m'$, and $C2'/c'$.
Solutions yielding a large uncompensated magnetization were excluded, as they conflict with the magnetometry measurements shown in Figs.~\ref{fig2}(c) and \ref{fig2}(d).
Crucially, the absence of the $001$ diffraction peak is decisive for our refinement; this reflection only gains magnetic intensity only if the N\'{e}el vector tilts away from the $c$ axis (see Appendix~\ref{app:neutrons} for details).
Based on the background level at this position, we conclude that any possible tilt cannot exceed $\sim 0.5^{\circ}$. 
Our neutron-diffraction results, therefore, confirm the previously reported $P6_{3}'/m'm'c$ symmetry with a propagation vector $k = (0,0,0)$, as shown in the inset of Fig.~\ref{fig1}(c). 
From the refinement, we obtain a sublattice magnetic moment of $(2.48 \pm 0.13) \, \mu_{\mathrm{B}}/\text{Cr}$, consistent with previous reports \cite{Yuan2020}.

\subsection{Magnetoresistance and Hall effect}

\begin{figure*}[tb]
    \centering
    \includegraphics[width=\textwidth]{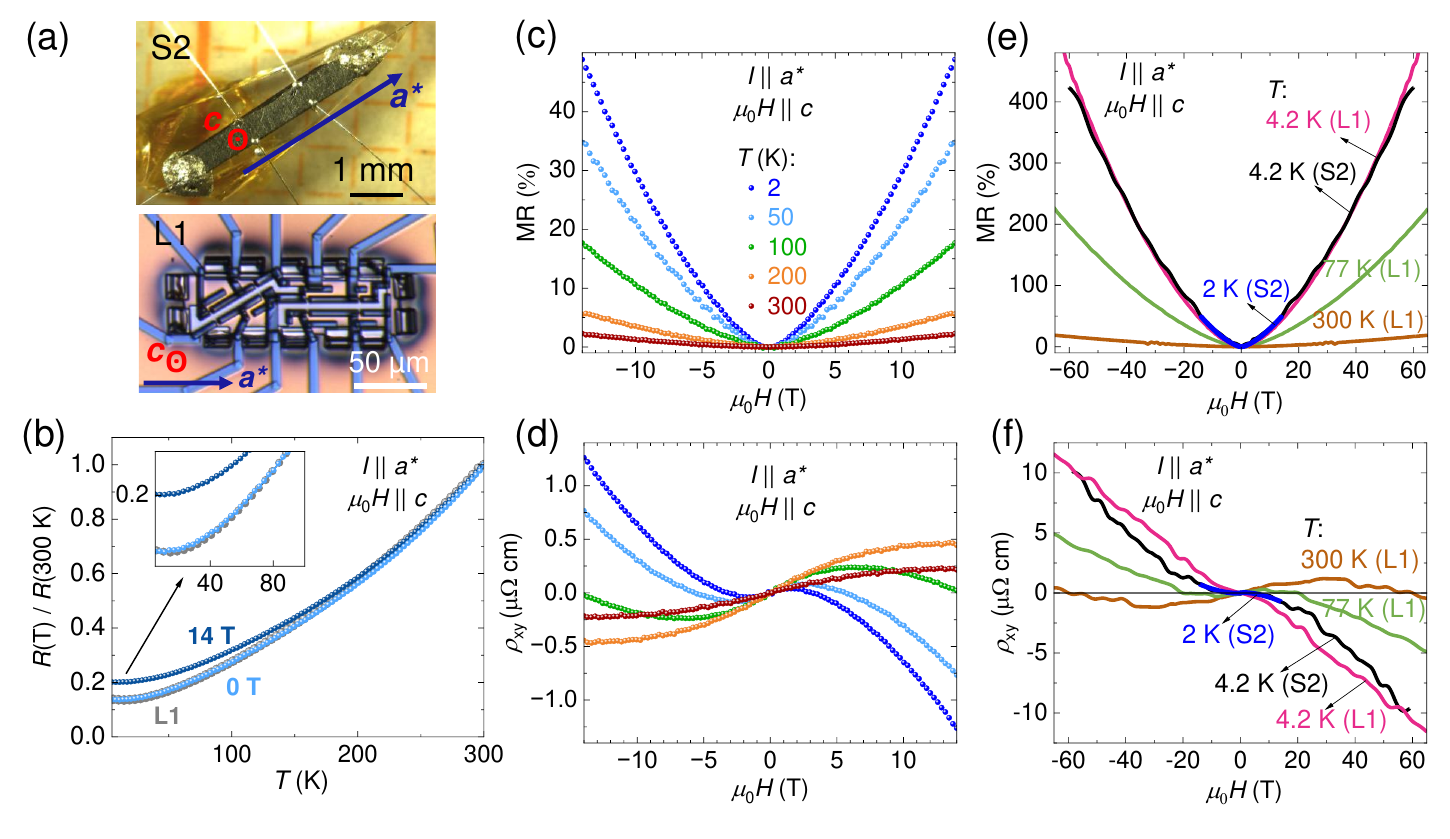}
    \caption{\textbf{Electrical transport measurements: (a)} Optical images of the Hall bar bulk sample S2 and the multiterminal microstructure L1.
    Current is applied along the $a^{*}$ axis, and magnetic field is applied along the $c$ axis for both samples. 
    \textbf{(b)} Longitudinal resistivity measured at 0 and 14\,T. Inset: enlargement of the low temperature resistivity. The zero-field resistivity of L1 (grey) matches well with that of the bulk sample (light blue). 
    \textbf{(c, d)} Magnetic-field dependence of the magnetoresistance and Hall resistivity of the sample S2 recorded at various fixed temperatures between 2 and \SI{300}{\kelvin}.
    \textbf{(e, f)} MR and Hall effect recorded in pulsed fields up to 65\,T. We compare steady-field data recorded at 2\,K (blue) and pulsed-field data at 4.2\,K (black) for S2 with pulsed-field data of sample L1 measured at 4.2, 77, and 300\,K, pink, green, brown, respectively.}
    \label{fig3}
\end{figure*}

We further studied the field-dependent electrical magnetotransport of a single-crystalline CrSb bar (S2) [Fig.~\ref{fig3}(a)] between room temperature and 5\,K. 
In Fig.~\ref{fig3}(b), we show the temperature dependence of the longitudinal resistivity $\rho_{xx}$, measured between 5 and 300 K for zero field and 14\,T, respectively.
The room-temperature resistivity $\rho_{xx}(300\,\mathrm{K})=$ \SI{52.7}{\micro\ohm\centi\meter} is consistent with recent reports where it ranges from \SI{57}{\micro\ohm\centi\meter}~\cite{peng2024} to \SI{70}{\micro\ohm\centi\meter}~\cite{bai2025} at \SI{300}{\kelvin}.
The residual resistivity ratio (RRR) of 7.3 indicates a good metallicity and minor structural or chemical defects.
We have reproduced this RRR for various single-crystalline samples and have consistently found values between 7 and 10. 
As can be seen in Fig.~\ref{fig3}(c), the magnetoresistance (MR), i.e., MR = $\rho_{xx}(B)/\rho_0-1$, exhibits a nonsaturating almost quadratic dependence with magnetic field.
Its magnitude increases with decreasing temperature, ranging from a few per cent at room temperature to more than $50\%$ at 2\,K and 14\,T.

MR is, however, decreasing with increased temperature. 
This behavior is characteristic for metals and semimetals, in which the MR effect is more pronounced at lower temperatures, because of reduced phonon scattering.

In Fig.~\ref{fig3}(d), we present Hall-effect data recorded simultaneously with the MR data.
$\rho_{xy}$ shows a nonlinear field dependence for all investigated temperatures, which gets more pronounced at lower temperature.
While for temperatures above 100\,K the Hall-resistivity slope remains positive for the full field range up to 14\,T, it crosses zero and turns negative at ever decreasing field values upon decreasing temperature. 
We extended the field range up to 65\,T by conducting magnetotransport measurements in pulsed magnetic fields for the two samples S2 and L1 [Fig.~\ref{fig3}(a)] with $\mu_0H\parallel c$, see Figs.~\ref{fig3}(e) and \ref{fig3}(f).
For both samples, the MR does not show any sign of saturation up to the highest fields.
Thanks to the reduced cross section of the microfabricated sample L1, the overall signal-to-noise ratio could be improved, as can be seen in the curves recorded at 4.2\,K for S2 (black) and L1 (pink) in Fig.~\ref{fig3}(e). 

Interestingly, we find that the Hall resistivity [Fig.~\ref{fig3}(f)] acquires an overall negative slope in the high-field range for all measured temperatures.
We also observe slight differences in the overall field dependence of the Hall effect between the two samples.
This may originate from differences in the purity or additional built-in stress induced in the microstructure from its fixation to the substrate.
However, the overall negative slope at high fields appears to be a general property of CrSb.

To further analyze the MR and Hall data we have performed a multicarrier analysis following previous reports~\cite{urata2024, peng2024, bai2025}.
We provide details in appendix~\ref{app:multicarrier}.
Since it is \textit{a priori} unclear how many bands need to be considered, we have performed the analysis with two, three, and four individual contributions.

\begin{figure}[tb]
    \centering
    \includegraphics[width=1\columnwidth]{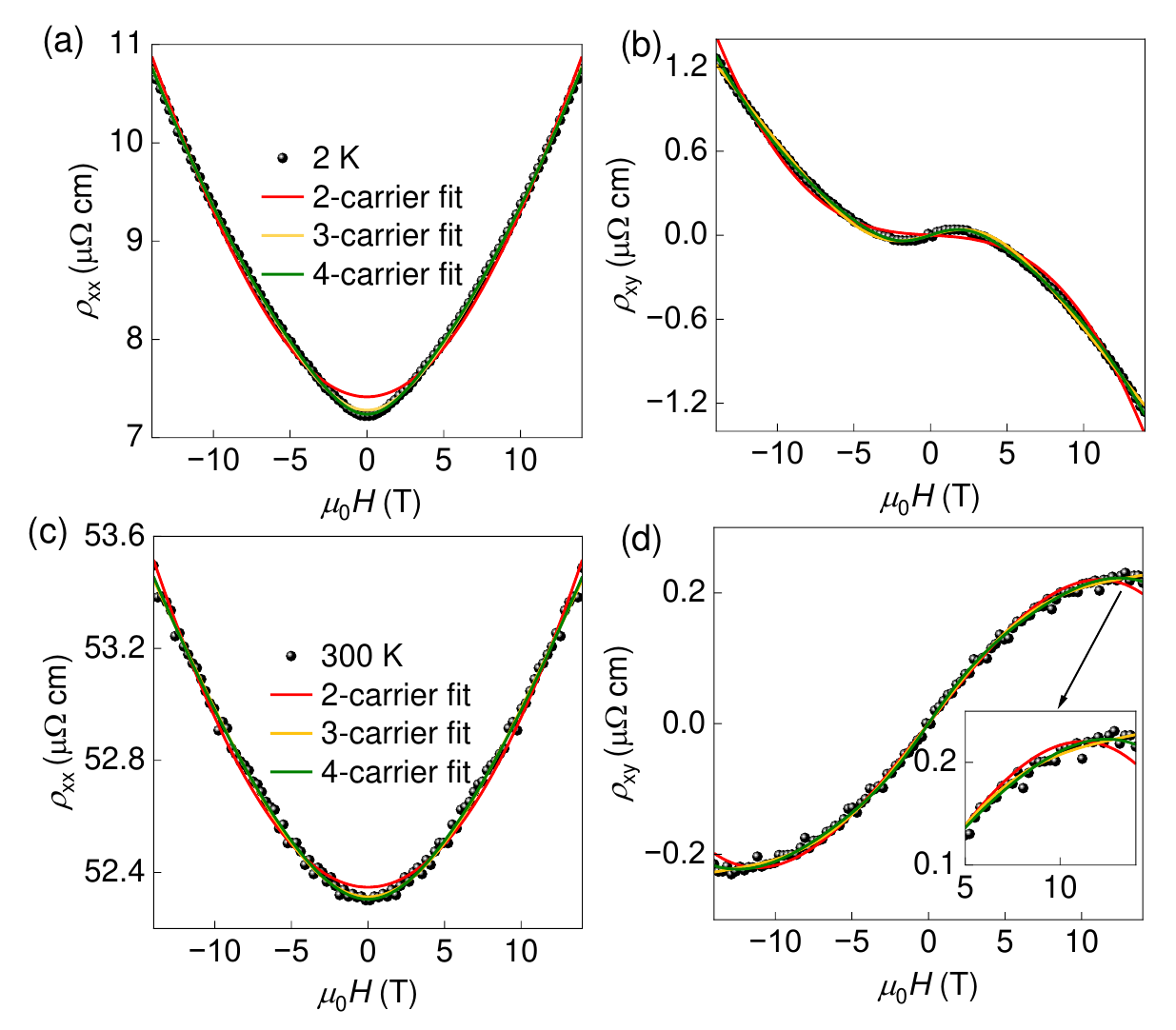}
    \caption{\textbf{Multicarrier fitting of MR and Hall resistivity:}
     Magnetic-field dependence of the longitudinal resistivity at 2 \textbf{(a, c)} and \SI{300}{\kelvin} \textbf{(b, d)}, respectively. 
    In all panels, the experimental data are compared to fits considering the contribution of 2, 3, and 4 individual types of charge carriers. Inset in (d): enlarged high-field range, highlighting the differences of the models.
    }
    \label{fig4}
\end{figure}

The fit gradually improves as more bands and, therefore, more parameters are added, as seen in Fig.~\ref{fig4}.
In order to obtain a reasonable agreement with the experimental data recorded at \SI{2}{\kelvin} and fields up to \SI{14}{\tesla} [Figs.~\ref{fig4}(a) and \ref{fig4}(b)], at least three bands are required. 
Adding a fourth band, the matching with the Hall resistivity at higher magnetic fields is improved as seen in the inset of Fig.~\ref{fig4}(d).
By contrast, for smaller magnetic fields up to \SI{10}{\tesla} at \SI{300}{\kelvin} already two bands reasonably well reproduce both MR and Hall data.
In the following, to describe the data up to \SI{14}{\tesla}, we therefore, chose four bands to perform a multicarrier analysis over the accessed temperature range between 2 and 300\,K.
Moreover, our recent work on the study of Shubnikov-de Haas oscillations and its analysis using the density-functional theory has also confirmed multibands in CrSb~\cite{Naduvile2026}.

\begin{figure}[tb]
    \centering
    \includegraphics[
        width=1\columnwidth]{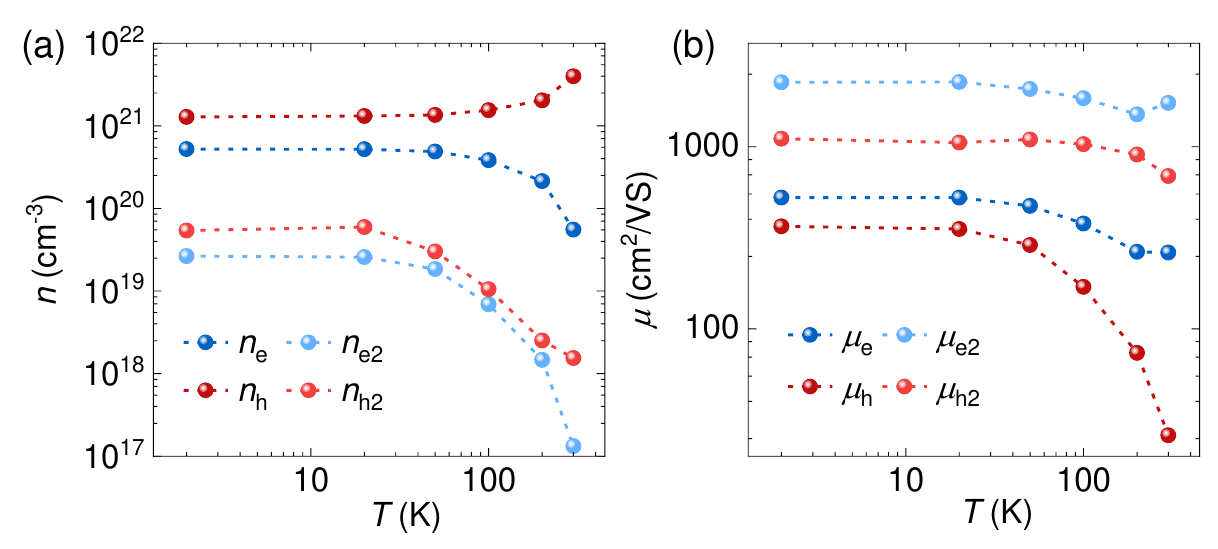}
    \caption{\textbf{Multicarrier analysis:}
    Temperature dependence of \textbf{(a)} the charge-carrier concentration and \textbf{(b)} mobility obtained from the four-band model fitting.}
    \label{fig5}
\end{figure}

The temperature dependence of the model parameters is shown in Fig.~\ref{fig5}.
The dominant band (the one with the highest charge carrier concentration) is of hole-like character, consistent with previous reports~\cite{urata2024, peng2024, bai2025}.
At temperatures below \SI{20}{\kelvin}, we find stable fit parameters and sizable mobilities of electron- and hole-like carriers.
At higher temperatures, in particular above \SI{100}{\kelvin}, the mobilities decrease.

\begin{figure}[tb]
    \centering
    \includegraphics[width=0.9\columnwidth]{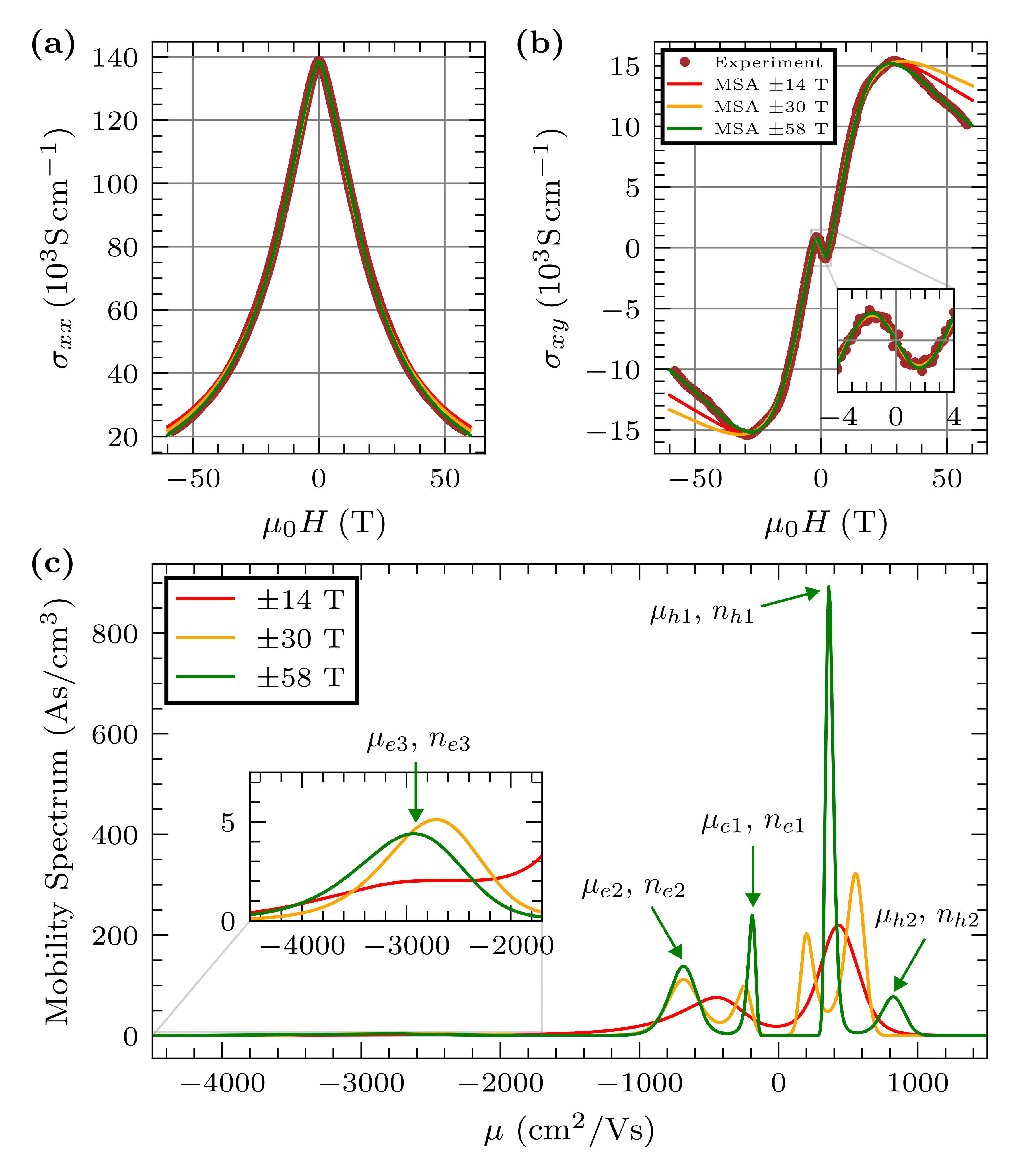}
    \caption{\textbf{Mobility spectrum analysis (a,b)} Conductivity reconstructed from the respective MSA models evaluated over different magnetic-field ranges ($\pm$14, $\pm$30, and $\pm$58\,T).
    \textbf{(c)} Mobility spectra obtained for various magnetic-field intervals.}
    \label{fig:msa}
\end{figure}

Additionally, we employ the mobility-spectrum-analysis (MSA) technique, which extracts information about the charge-carrier distribution in a material from magnetotransport measurements~\cite{Beck1987, Beck2021}.
MSA decomposes the measured longitudinal- and Hall-resistivity data sets into continuous spectra of carrier contributions as a function of mobility, meaning $\sigma_{i} = n_iq\mu_i \rightarrow \sigma(\mu)$~\cite{Beck1987}.
This allows for the simultaneous determination of the sign (electron- or hole-like), the relative weight (contribution to conductivity), and the corresponding mobility without any assumption on the number of contributing charge carrier types. A more detailed description of the MSA is provided in appendix~\ref{app:msa}.
In practice, the method provides a way to analyze complex transport data, especially in multiband systems. 
The method benefits from using as large a field range as possible for the analysis.
In Fig.~\ref{fig:msa} we present the modelling based on MSA for the 4.2\,K data recorded for sample S2 in three different field ranges.
For this purpose, the resistivity data [Fig.~\ref{fig3}] were converted to conductivity.
For increased sensitivity at lower fields, we have merged the pulsed-field data with data recorded in steady fields up to 14\,T.
When considering the full field range of \SI{-58}{\tesla} to \SI{58}{\tesla}, the resulting mobility spectrum, $\sigma(\mu)$, reveals five distinct peaks [Fig.~\ref{fig:msa}(c)]. Each peak is associated with a charge-carrier population contributing to the magnetic-field dependence of both longitudinal and transverse resistivities.
The sign of the mobility at each peak identifies the charge-carrier type, with negative values corresponding to electrons and positive values to holes.
The integrated spectral weight of each peak serves as a quantitative measure of its contribution to the total conductivity. This, in turn, provides insight into the associated charge-carrier density and enables a comparison of the relative contributions of different charge-carrier populations to the overall transport response.

\begin{table}
  \centering
  \caption{Parameters of charge-carrier populations (e - electrons and h - holes) from MSA  
  for sample S2 at 4.2\,K.
  Contributions of electrons are labeled e$i$ whereas hole like populations are labeled h$i$, with integer $i$ consistent with labels in Fig.~\ref{fig:msa}.}
  \label{tab:msa_5band}
  \begin{tabular}{lcc} 
    \hline
    & \multicolumn{2}{c}{MSA} 
    \\
    Label & $n$ ($\mathrm{cm^{-3}}$) & $\mu$ (cm$^{2}$/Vs)
            
            \\
    \hline
    e1 & $8.25 \times 10^{20}$ & 190 \\
    e2 & $4.31 \times 10^{20}$ & 680 \\ 
    e3 & $2.0 \times 10^{19}$  & 2930 \\ 
    h1 & $1.13 \times 10^{21}$ & 360 \\ 
    h2 & $1.16 \times 10^{20}$ & 820 \\ 
    \hline
    \label{tab:msa}
  \end{tabular}
\end{table}

Table~\ref{tab:msa} provides an overview of the identified charge-carrier populations, which overall compare well with the parameters found by the multi-carrier analysis [Fig.~\ref{fig5}].
The conductivity reconstructed from the mobility spectrum is shown in comparison with experimental data in Figs.~\ref{fig:msa}(a) and \ref{fig:msa}(b).
The presence of high-mobility bands points at predicted topological Weyl nodes in the altermagnetic band structure of CrSb~\cite{Li2025,Lu2025}.
The experimentally observed non-saturating MR may therefore be associated with the topologically nontrivial properties of CrSb's electronic band structure.

The MSA shows, consistent with previous studies, that multiple charge carriers contribute to the transport properties of CrSb.
In particular, as for our analysis with a fixed number of bands, we find that the dominating band with the highest charge-carrier concentration exhibits a hole-like character.
The various charge-carrier populations show sizable mobilities in the range up to $\sim 3000$ cm$^2$/Vs.
The exact number of charge carriers needed to describe the transport data differs in various studies~\cite{bai2025, urata2024, peng2024}.
These studies, however, also used different magnetic-field ranges.
Our analysis based on the MSA, using various magnetic field ranges shows that indeed the field range determines how many independent charge-carrier populations can be resolved (Fig.~\ref{fig:msa}).
For magnetic-field ranges from $\SI{\pm14}{\tesla}$ to $\SI{\pm58}{\tesla}$, we find that the number of identified charge-carrier populations increases from 3 to 5, covering the 3 or 4 charge carrier types reported in Refs.~\cite{bai2025, urata2024, peng2024}.
A hole-like band remains dominant band for all magnetic-field ranges, which is consistent with the overall Hall slope in the high-field range observed for the full temperature range between 2 and 300\,K.
When extending the magnetic-field range, the MSA can resolve more and more individual contributions of different charge-carrier types. 
Therefore, these results highlight the importance of utilizing an extended magnetic-field range for a reliable characterization and resolution of multicarrier transport. 

\subsection{Thermal conductivity and thermal Hall effect}

\begin{figure*}[tb]
    \centering
    \includegraphics[
        width=1\textwidth] {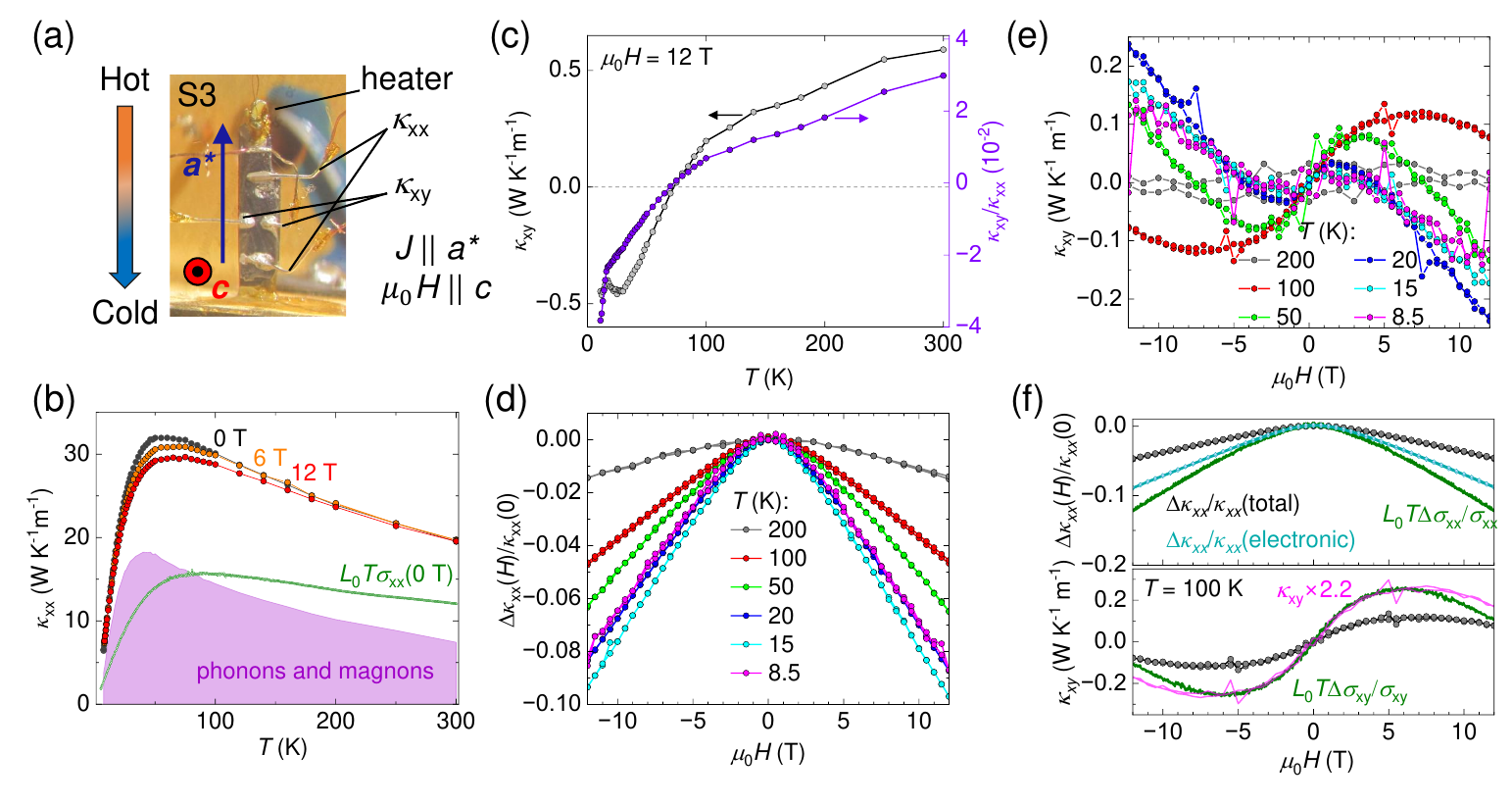}
    \caption{
        \textbf{Thermal transport measurements:}
        \textbf{(a)} Optical image of the bar-shaped sample S3.
        \textbf{(b)} Temperature dependence of the longitudinal thermal conductivity $\kappa_{xx}$ at selected magnetic fields. 
        The green curve indicates the electronic contribution to the thermal conductivity, calculated from the Wiedemann-Franz law.
        \textbf{(c)} Temperature dependence of the thermal Hall conductivity and the thermal Hall angle.
        \textbf{(d, e)} Magnetic-field dependence of the relative change in the thermal conductivity and the thermal Hall conductivity at selected temperatures. 
        \textbf{(f)} Comparison of field dependences of the thermal and thermal-Hall conductivity in the context of the Wiedemann--Franz law at 100\,K.
    }
    \label{fig:thermal}
\end{figure*}

We also carried out thermal-conductivity and thermal-Hall-effect measurements on two bulk CrSb single-crystals.
Figure~\ref{fig:thermal} presents an overview of the thermal-transport measurements on sample S3 performed between 7 and 300\,K and for magnetic fields up to 12\,T. 
Figure~\ref{fig:thermal}(a) shows a picture of S3 installed for the simultaneous measurement of thermal conductivity ($\kappa_\mathrm{xx}$) and thermal Hall effect ($\kappa_\mathrm{xy}$). 
A thermal gradient is applied along the long side of the sample within the hexagonal plane ($J\parallel a^{*}$), while the magnetic field is applied out of the plane ($\mu_0H\parallel c$).

In Fig.~\ref{fig:thermal}(b), we show the temperature dependence of $\kappa_{xx}$ at selected magnetic fields. 
The thermal conductivity increases rapidly towards 50\,K, then peaks, and gradually decreases toward room temperature. 
In the presence of a magnetic field, $\kappa_{xx}$ is slightly reduced, consistent with an additional magnetic-field-dependent scattering contribution. 
For comparison, we show the electronic contribution to the heat conduction, calculated from the Wiedemann-Franz law $L_0\sigma_{xx}T$~\cite{Wiedemann1853, Chester1961, Ott198} (green curve).
Our result show a nonelectronic contribution to the thermal conductivity, caused by phonons and magnons, as the difference between the measured and Wiedemann–Franz electronic contribution.
The estimated nonelectronic thermal conductivity peaks around 40\,K.
However, it does not decrease to zero at high temperatures, as expected from the $1/T$ law, which is typical for a magnonic contribution. 
In Fig.~\ref{fig:thermal}(c), we show the temperature dependence of the thermal-Hall conductivity and the thermal-Hall angle, $\theta_H = \kappa_{xy}/\kappa_{xx}$, measured in a $\pm 12\,$T out-of-plane field. 
The thermal Hall conductivity changes sign around 70\,K, suggesting competing contributions to the thermal Hall response.

In Fig.~\ref{fig:thermal}(d) and \ref{fig:thermal}(e), we show the magnetic-field dependence of the relative change in the thermal conductivity, $\Delta \kappa_{xx}(H) / \kappa_{xx}(0)$, and of $\kappa_{xy}(H)$ at selected temperatures, respectively. 
The magnetic-field dependence of the thermal conductivity is quadratic and increases in magnitude as the temperature decreases. 
The strongest field dependence is reached around 15\,K. 
The isothermal thermal Hall conductivity shows a nonlinear field dependence. 
At all temperatures, in the low-field region, the thermal Hall conductivity has a positive slope. 
However, below 70\,K, a large component with a negative slope dominates the high-field region.

Finally, to provide a better insight into the possible origin of the field dependence of the heat conduction in CrSb, we compare the relative thermal-conductivity with the electronic component calculated from the magnetoresistance measurements (green) at 100\,K, as shown in Fig.~\ref{fig:thermal}(f). 
We can calculate the corresponding $\Delta \kappa / \kappa$ ratio using different base values: the change can be related either to the total thermal conductivity, $\Delta \kappa_{xx} / \kappa_{xx}$((total) (black), or to the electronic component, $\Delta \kappa_{xx} / \kappa_{xx}$(electronic) (cyan). 
As shown in Fig.~\ref{fig:thermal}(f), top panel, the ratio $\Delta \kappa_{xx} / \kappa_{xx}$(total) (black curve) deviates substantially from the electronic component calculated from the magnetoresistance measurements (green), whereas $\Delta \kappa_{xx} / \kappa_{xx}$(electronic) (cyan) matches much better, though not perfectly. 
This suggests that, while the field dependence of the heat conduction is predominantly dominated by the electrons, a minor contribution may be related to phonons~\cite{Debye1912} and magnons~\cite{Kittel1951, Rezende2019}. 
In the lower panel of Fig.~\ref{fig:thermal}(f), the thermal Hall conductivity (black) deviates from that calculated from the electronic contribution via the Wiedemann-Franz law (green). 
The total and electronic contributions can be brought onto a similar scale by an empirical factor of 2.2 (magenta); however, at higher fields, still deviations appear.

Taken together, our results establish a consistent picture of CrSb as a compensated altermagnetic semimetal with distinct multicarrier transport.
On the electronic side, the extended magnetic-field range is essential for resolving the individual transport channels and reconciling the different charge carrier counts reported previously.
On the thermal side, the nonlinear thermal Hall response, its sign change around 70\,K, and the comparison to the Wiedemann-Franz law show that the heat transport follows the electronic response in broad terms, yet cannot be described fully within a single conventional metallic contribution.
We, therefore, conclude that electrons provide the dominant contribution to both charge transport and the field-dependent part of the heat transport in CrSb, while additional heat-transport channels, caused by phononic and magnonic contributions, are relevant for the thermal response.
In this sense, CrSb provides a useful benchmark system in which multicarrier semimetallic transport and thermal Hall physics can be addressed within a single material platform with an experimentally well-established altermagnetic state.

\section{Conclusion}

We have synthesized high-quality CrSb single crystals using chemical vapor transport and conducted a comprehensive systematic study of the electrical and thermal magnetotransport properties.
We observed a distinct N\'eel temperature ($T_\mathrm{N}\approx700$\, K) in resistivity and magnetization measurements. 
Magnetization and neutron diffraction confirm the compensated magnetic ordering in CrSb.
Notably, CrSb single crystals exhibit large, positive, and nonsaturating magnetoresistance.
Another intriguing observation is the nonlinear Hall effect at temperatures ranging from \SI{2}{\kelvin} to \SI{300}{\kelvin}, which we successfully describe by a four-carrier Drude model.
Moreover, our detailed mobility spectrum analysis, taking high-field data up to \SI{58}{\tesla} into account, confirms the presence of very-high-mobility bands ($\mu_{\mathrm{h}2}\approx800$ and $\mu_{\mathrm{e}3}\approx3000$\,cm$^2/$Vs) that contribute to the electrical transport at low temperature.
This provides experimental evidence of the topologically non-trivial properties of the altermagnetic band structure in CrSb.
These findings suggest that the electronic band structure and Fermi-surface topology play a significant role in shaping the electrical transport properties of CrSb.
The overall field and temperature dependence of the thermal conductivity and thermal Hall response broadly follows that observed in the electrical transport.
The thermal conductivity appears to be significantly enhanced relative to the simple Wiedemann–Franz law.
This significant deviation at low temperatures further indicates that heat transport is not solely governed by the electronic charge carriers, suggesting a substantial contribution from magnons or phonons in this $g$-wave altermagnetic system.
Furthermore, the sensitivity of the multicarrier fit to the magnetic field range highlights the complex topology of the Fermi surface in CrSb.
Our results identify CrSb as a benchmark altermagnetic semimetal and show that high-field transport is essential for disentangling multicarrier electrical and thermal responses in this class of materials.
CrSb bears the potential as premier platform for studies of the interplay between altermagnetism, high-mobility topological fermions, and anomalous thermal transport.

\section{Acknowledgments}

We gratefully acknowledge the contributions of Dr. Karel Výborný (deceased) to the development of the multi-carrier analysis framework employed in this study.
We acknowledge funding from the Czech Science Foundation (Grant No. 22-22000M), and Lumina Quaeruntur fellowship LQ100102201 of the Czech Academy of Sciences.
Single crystals were grown and characterized in MGML, which is supported within the program of Czech Research Infrastructures (project No. LM2023065).
HR was supported by TERAFIT - CZ.02.01.01/00/22$\_$008/0004594 and the Max Dioscuri Program LV23025 funded by MPG and MEYS.
Access to the Ion Beam Center (IBC) of the HZDR is gratefully acknowledged.
We acknowledge support from the Deutsche Forschungsgemeinschaft (DFG) through Grant No. HE 8556/3-1 and the
W\"{u}rzburg-Dresden Cluster of Excellence on Complexity, Topology and Dynamics in Quantum Matter--$ctd.qmat$ (EXC 2147, Project No.\ 390858490), as well as the support of the HLD at HZDR, member of the European Magnetic Field Laboratory (EMFL).
We acknowledge financial support from DFG through the Individual Research Grant (project-id 540912241), the Collaborative Research Center SFB 1143 (project-id 247310070).
Part of this work was funded by the European Union as part of the Horizon Europe call HORIZON-INFRA-2021-SERV-01 under grant agreement number 101058414 and co-funded by UK Research and Innovation (UKRI) under the UK government’s Horizon Europe funding guarantee (grant number 10039728) and by the Swiss State Secretariat for Education, Research and Innovation (SERI) under contract number 22.00187.
Views and opinions expressed are however those of the author(s) only and do not necessarily reflect those of the European Union or the UK Science and Technology Facilities Council or the Swiss State Secretariat for Education, Research and Innovation (SERI).
Neither the European Union nor the granting authorities can be held responsible for them.
The authors acknowledge the Science and Technology facility council (STFC) for the provision on neutron beam time on the WISH diffractometer at the ISIS facility under the proposal RB2510416.
Raw data of the neutron measurements can be found at~\cite{kriegnerrb2510416}.

\appendix

\section{Neutron-diffraction analysis}
\label{app:neutrons}

\begin{figure}[tbh]
    \centering
    \includegraphics[
       width=0.9\columnwidth]{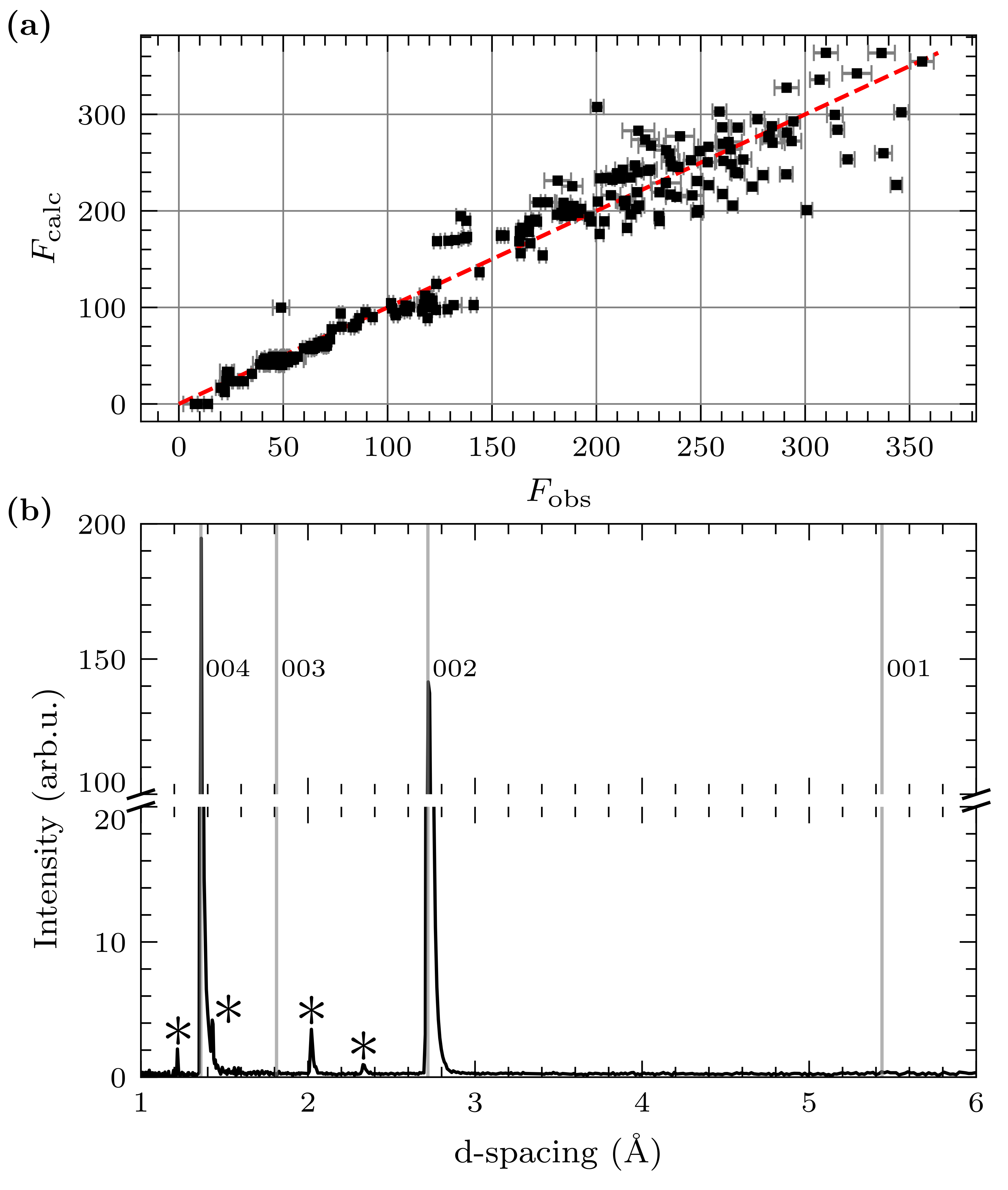}
    \vspace{3mm}
    \caption{\textbf{Neutron diffraction correlation plot and $00l$ line cut:}
    \textbf{(a)} Correlation plot of observed versus calculated structure factors. Multiple observations of the same diffraction peak for different wavelengths and sample rotation are shown as individual data points. The associated statistical uncertainties of the experimental data are shown as error bars.
    \textbf{(b)} Measured intensity along the $00l$ direction showing the absence of peaks with odd $l$. The signal from Al (indicated by $\ast$) originates from the sample holder.}
    \label{fig:neutrons}
\end{figure}

The correlation plot of the calculated versus observed structure factor for the single-crystal WISH data is shown in Fig.~\ref{fig:neutrons}(a).
The pronounced linear correlation between $F_\text{obs}$ and $F_\text{calc}$ demonstrates that the refined structural model satisfactorily reproduces the measured intensities.
With respect to an eventual tilt of the N\'eel vector a particularly important observation is the absence of the $00l$ diffraction peaks with $l=1,3,\dots$ (odd).
This is shown in Fig.~\ref{fig:neutrons}(b) where exclusively signal for CrSb $00l$ for $l=2,4$. 

\section{Longitudinal and transverse resistivity}

We symmetrized the measured longitudinal resistivity ($\rho_{xx}$) and anti-symmetrized the Hall resistivity data ($\rho_{xy}$) with respect to the external applied magnetic field using the equations:

\begin{equation}
\rho_{xx}(\mu_0H) = \frac{\rho_{xx}(+\mu_0H) + \rho_{xx}(-\mu_0H)}{2},
\label{symmetrisation}
\end{equation}

\begin{equation}
\rho_{xy}(\mu_0H) = \frac{\rho_{xy}(+\mu_0H) - \rho_{xy}(-\mu_0H)}{2}.
\label{antisymmetrisation}
\end{equation}

With this approach, we effectively eliminate additional contributions and artifacts introduced from geometric irregularities in the sample's shape or the positioning of the contacts.

We used a digital lock-in technique to process the pulsed-field data. We applied a third-order filter with a time constant of \SI{300}{\micro\second} for smoothing the measured data. We further smoothed data using a fast Fourier transform method in order to reduce noise during the magnetic pulses.

\section{Multicarrier fitting}
\label{app:multicarrier}

For the multicarrier analysis with the magnetic field applied along the $z$ direction ($\mu_0H \parallel z$), the conductivity tensor of the $i$-th charge carrier type within the classical Drude model can be written as

\begin{align}
    \sigma_i = \dfrac{n_i e \mu_i}{1 + (\mu_i \mu_0H)^2} \begin{pmatrix}
1 & s_i\mu_i \mu_0H  \\
-s_i\mu_i \mu_0H & 1 
\end{pmatrix},
\label{drude}
\end{align}
where $i$ denotes the carrier index, $e$ is the elementary charge, $n_i$ the carrier density, $\mu_i$ the carrier mobility, and $s_i$ the sign of the charge carrier with $s_i = -1$ for electrons and $s_i = +1$ for holes \cite{Kim1993}. The charge carrier densities $n_i$ and mobilities $\mu_i$ were treated as free parameters in the fitting procedure. For a system with $N$ carrier types, the total conductivity tensor is obtained by summing over all individual contributions, yielding

\begin{align}
    \sigma_{xx}(\mu_0H) &= \sum_{i=1}^{N} 
    \frac{n_i e \mu_i}{1 + (\mu_i \mu_0H)^2},
    \label{eq:sigma_xx} \\
    \sigma_{xy}(\mu_0H) &= \sum_{i=1}^{N} 
    \frac{n_i e \mu_i^2 \mu_0H}{1 + (\mu_i \mu_0H)^2} \, s_i.
    \label{eq:sigma_xy}
\end{align}

\begin{figure}[tbh]
    \centering
    \includegraphics[
        width=0.9\columnwidth]{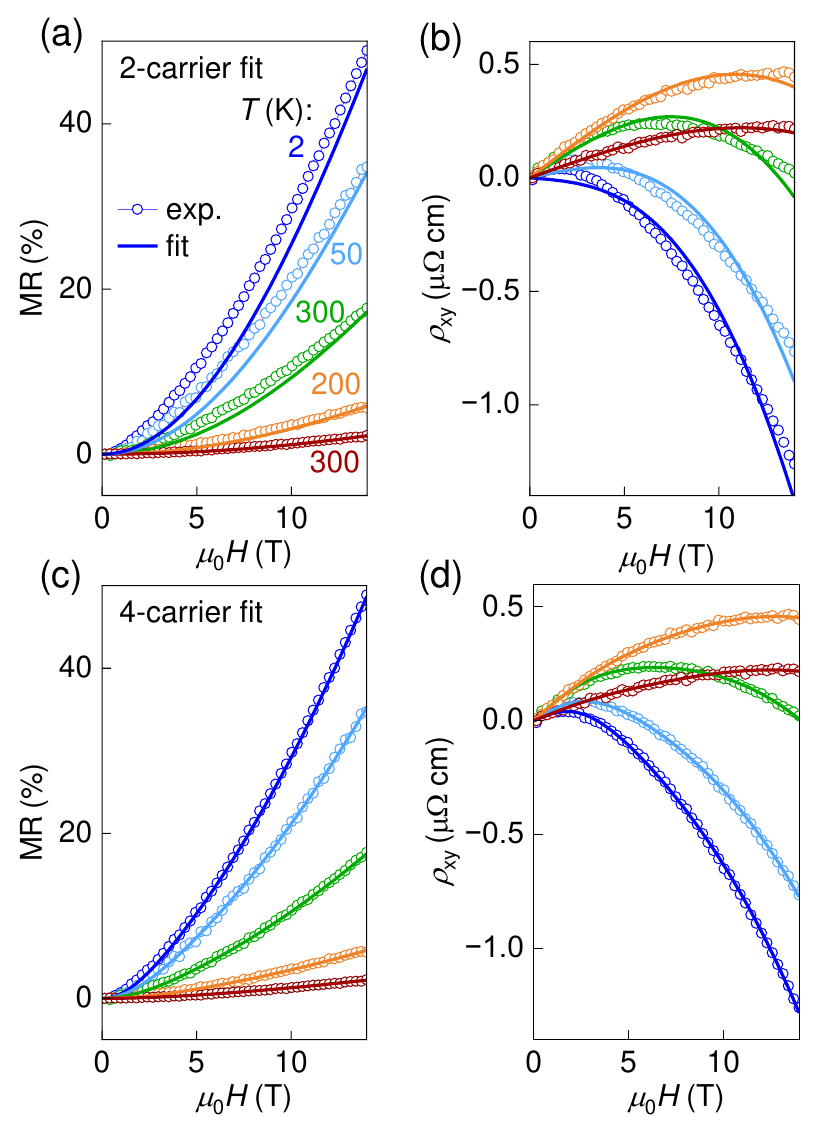}

    \caption{\textbf{Multicarrier fitting of MR and Hall resistivity: (a, b)}
    Magnetic-field dependence of the longitudinal resistivity and Hall resistivity compared to fits of a two-band model.
    \textbf{(c, d)} Magnetic field dependence of the longitudinal resistivity and Hall resistivity, respectively are compared to a four-band model.
    $\rho_{xx}$ is normalized to the zero-field resistivity of the respective data.
    Lines correspond to the respective fits.}
    \label{fig:MCAfit}
\end{figure}

Representative fits obtained within this multicarrier framework are shown in Fig.~\ref{fig:MCAfit} for models including two and four charge carrier types, respectively.

\section{Mobility spectrum analysis (MSA)}
\label{app:msa}
The dataset employed for the MSA combines steady-field data recorded up to $\pm$14\,T with pulsed-field data for $|\mu_0H|>14$\,T. 
This approach was necessary because the higher noise level in the pulsed-field measurements obscures key features in the low-field regime, most notably the slope change at approximately $\pm 2.5$\,T.

The concept underlying the mobility spectrum analysis is that, in the limit where the number of charge-carrier types $N$ in Eqs.~(\ref{eq:sigma_xx}) and (\ref{eq:sigma_xy}) becomes very large ($N \rightarrow \infty$), the discrete sum over N can be replaced by a continuous integral over the mobility $\mu$. In this formalism, the conductivity contribution of individual charge-carrier types is represented by a continuous conductivity spectrum $\sigma(\mu)$. For a magnetic field applied along the $z$ direction, the longitudinal and transverse components of the conductivity tensor are then given by

\begin{equation}
    \sigma_{xx}(\mu_0H) =
    \int_{-\infty}^{\infty}
    \frac{\sigma(\mu)}{1+(\mu \mu_0H)^2}\, d\mu,
\end{equation}

and

\begin{equation}
    \sigma_{xy}(\mu_0H) =
    \int_{-\infty}^{\infty}
    \frac{\mu \mu_0H\, \sigma(\mu)}{1+(\mu \mu_0H)^2}\, d\mu.
\end{equation}

In contrast to the discrete multicarrier model, where the sign of the charge carrier is explicitly introduced via the factor $s_i$, the MSA incorporates the charge-carrier type through the sign of the mobility. Negative and positive mobilities correspond to electron- and hole-like carriers, respectively, while the conductivity spectrum $\sigma(\mu)$ is required to be positive, $\sigma(\mu) \ge 0$, to ensure a physically meaningful representation.

\section{FIB-assisted fabrication of microstructures}
\label{app:FIB}

We fabricated CrSb microstructures using gallium focused-ion-beam (FIB) lithography. 
First, we cut a lamella with dimensions of the order of$(30 \times 2.5 \times 150)$\SI{}{\micro\meter}$^3$ from the bulk single crystal using a Ga-FIB and subsequently transferred it ex-situ onto a sapphire substrate. 
Next, an approximately \SI{150}{\nano\meter}-thick gold layer was deposited over the entire lamella via magnetron sputtering. 
The gold layer was then partially removed from the central top surface using Ga$^{+}$ ion etching. 
Subsequently, trenches were milled into both the gold layer and the underlying sample using FIB to define well-controlled contact terminals.
We patterned multiple voltage leads along the bar to enable measurements of both longitudinal and transverse voltages.
Further details about the procedure is well explained in Ref.~\cite{Moritz2022}.
Microstructure L1 [shown in Fig.~\ref{fig3}(a)] with the long axis parallel to the $a^{*}$ axis has a cross-section of $(w \times t) = (4.3 \times 2.1)$\SI{}{\micro\meter}$^2$ and a length of \SI{32.1}{\micro\meter} between the Voltage contacts.
We performed magnetotransport measurements with $I\parallel a^{*}$ and $\mu_0H\parallel c$.
In Fig.~\ref{fig3}(b), we present the temperature dependence of the zero-field longitudinal resistivity $\rho_{xx}$ of L1, measured between 5 and 300~K.
The room-temperature resistivity is \SI{64.4}{\micro\ohm\centi\meter} and the residual resistivity ratio is about 7.
Both values are consistent with previously reported data and with those expected for bulk CrSb.

\bibliography{Bibliography}

\end{document}